\documentstyle[12pt]{article}
\begin{document}
     \title{Next-to-Leading Order QCD Formulation of Deep Inelastic
               Scattering}
     \author{M. A. G. Aivazis\rlap,{${}^{a}$}
             Wu-Ki Tung\rlap,{${}^{a,b}$}  and
             Fredrick Olness\rlap,{${}^{c}$}\footnote[3]{
     This work is supported by the U.S. Department of Energy and the
    National Science Foundation. Paper presented by  Fredrick Olness. }\\
     \vspace{0.1cm}
     {\em ${}^{a}$Illinois Institute of Technology, Chicago, IL  60616 }\\
     \vspace{0.1cm}
     {\em ${}^{b}$Fermilab National Accelerator Laboratory, Batavia, IL  }\\
     \vspace{0.1cm}
     {\em ${}^{c}$Southern Methodist University, Dallas, Texas 75275  }\\
      }
     \maketitle

\begin{abstract}

We have performed a QCD next-to-leading order (NLO) calculation for Deep
Inelastic Scattering (DIS) retaining the full parton and hadron mass
dependencies. We find that the gluon initiated contributions to DIS
processes, such as charm production, are {\it comparable} in magnitude
({\it i.e.}, $30\%$ to $100\%$) to the ``leading-order'' (LO)
sea-quark  processes. The ``slow-rescaling" prescription and the full
NLO formalism are compared in a quantitative manner.  The use of DIS
distributions and the inclusion of the charm mass via slow-rescaling
are not sufficient to mimic the correct NLO physics.  These results
imply that previous analyses of charm production data to extract the
strange and charm content of the nucleon, as well as the precise
determination of Standard Model parameters based on these analyses
(such as the Weinberg angle), need to be reassessed.

\end{abstract}

Total inclusive deep inelastic scattering of electrons, muons, and
neutrinos on nucleons is  the primary source of information on parton
distributions. Global analysis of the total  inclusive data  does not,
however, provide a good handle on the strange and charm quark  content
of the nucleon.  Specifically, it was emphasized  at the 1989
Breckenridge meeting and the  1990 Snowmass  meeting that
uncertainties of the strange-quark distribution  represent a large
source of error in Standard Model phenomenology; in the determination
of the Weinberg angle, it  is the dominant error.  With new
high-precision experiments now generating data\rlap,\footnote{ {\it
C.f.} presentations  by M.~Shaevitz and W.~G.~Cobau  in these
proceedings.}
 it is imperative  to  advance the theoretical calculations for these
experiments.  Since perturbative Quantum Chromodynamics  (QCD) provides
a  comprehensive framework to describe these processes, it is clearly
evident that a  proper analysis of deep inelastic scattering must be
carried out to NLO  in  QCD.

We have performed a QCD NLO calculation for Deep Inelastic Scattering
(DIS) retaining the full parton and hadron mass dependencies.\footnote{
See M. Aivazis, F. Olness, and W. Tung,  {\it Phys. Rev. Lett.} {\bf
65}, (1990) 2339, and references contained therein. } We find that the
NLO processes can  be comparable in magnitude to the leading  order
ones.   Specifically, the gluon initiated contributions to DIS
processes, such as charm production, are {\it comparable} in magnitude
({\it i.e.}, $30\%$ to $100\%$) to the LO sea-quark  processes,  ({\it
c.f.}, Fig.~1). Clearly, the (apparently) higher-order  contributions
cannot be neglected, and a NLO calculation is necessary to
unambiguously define the  strange and gluon parton distributions.
These results imply that previous analyses  of  charm production data
to extract the strange and charm content of  the nucleon, as well as
the  precise determination of Standard Model  parameters based on these
analyses, need to be  reassessed.

To establish the implications of this calculation for the determination
of the parton distributions,  these NLO results must be systematically
compared with the existing  LO calculation used to extract the PDF's
from the experimental structure functions. The conventional method
assumes the lowest order  process   $W+s,d \rightarrow c$ with zero
quark masses, and introduces the charm mass dependence only via  the
``slow-rescaling'' prescription:   $x_{bj} \rightarrow \xi_c = x_{bj}(1
+ m_c^2/Q^2) $. The effect of rescaling is shown in  Figure~2. We plot
$d\sigma/dy$ for the process  $W+s  \rightarrow c$ with an incident
neutrino energy of  $E_{\nu}=80\ GeV$, a characteristic value for the
fixed-target  DIS experiments.

As is well-known,  the ``slow-rescaling'' prescription  {\it
dramatically} reduces the cross section---by an order-of-magnitude for
low-$y$. This means that we must be very careful about: {\it i)} what
value we use for the charm quark mass,   {\it ii)} and, more
importantly,  how we introduce the charm quark mass dependence into
the  formulas.  We note that even in LO, the hard (Born) matrix element
has explicit charm mass dependence which is always neglected in the
conventional analysis. This is clearly  unacceptable.

To illustrate the charm quark mass dependence of the cross section,  we
plot $d\sigma / dy$ (using the complete NLO calculation)  in Figure~3
for a range of mass values. Because  $d\sigma / dy$  is very sensitive
to the charm mass,  an accurate measurement cross section measurement
has the potential  to precisely  determine the charm quark mass.
Conversely, an imprecise calculation (such as one lacking  the proper
mass dependence and the NLO corrections)  can be forced to agree with
data by shifting the input value of the charm mass by a small amount.
In particular, only with the inclusion of NLO terms does the charm
quark mass acquire an unambiguous meaning in perturbative  QCD.

We  compute both the LO and NLO processes retaining the full mass
dependence. The results from this calculation are shown in Figure~1.
The LO contribution  is computed from the Born diagram, but with the
quark masses retained. The scaling variable $\xi_c$  emerges naturally;
no ``slow-rescaling'' prescription is needed. The magnitude of the NLO
gluon contribution is displayed. In this case, the NLO contribution is
negative. The complete NLO term includes the $2\rightarrow 2$
sub-process $W+g\rightarrow q + \bar{q}$ as well as the ``subtraction"
term. (Only the combination is  well-defined.) The total contribution
(LO+NLO)  lies below the LO curve. Note that the NLO distribution is
relatively independent of $y$, whereas the LO falls off steeply for
small $y$.  This means that the NLO contribution will become very
significant at small $y$. In particular, it will affect the  slope and
$y$-intercept of the differential cross section.

It is now very interesting to compare the  ``slow-rescaling''
prescription to the complete NLO calculation with the full mass
dependence retained. Comparing Figs.~1 and~2, we see that the
``slow-rescaling'' prescription
 has succeeded in bringing the bulk of the cross section  into line
with the full NLO calculation. To make a more precise comparison, we
plot the ratio of the cross sections in Figure~4. We find that there
are two important deficiencies with the ``slow-rescaling'' prescription:
{\it 1)} The absolute normalization is not the same, and  {\it 2)} the
shape of the differential cross section is not the same.  Therefore,
we conclude that the ``slow-rescaling'' prescription  may be  sufficient
to grossly approximate the complete mass dependence at the level of  a
factor of 2; but, it is not sufficient for high  precision calculation
and measurements.

\end{document}